# Electron transporting bilayers for perovskite solar cells: spray coating deposition of c-TiO$_2$/m-SnO$_2$-quantum dots

**Anca G. Mirea[1†], Ioana D. Vlaicu[1†], Sarah Derbali[1], Florentina Neatu[1], Andrei G. Tomulescu[1], Cristina Besleaga[1], Monica Enculescu[1], Andrei C. Kuncser[1], Alexandra C. Iacoban[1], Nicolae Filipoiu[2,3], Marina Cuzminschi[2,3], George A. Nemnes[2,3], Andrei Manolescu,[4] Mihaela Florea,[1*] Ioana Pintilie[1*]**

[1] National Institute of Materials Physics, 405A Atomistilor Str., Bucharest, 077125, Romania

[2] Faculty of Physics, University of Bucharest, Magurele 077125, Ilfov, Romania

[3] Horia Hulubei National Institute of Physics and Nuclear Engineering, 30 Reactorului Str., Magurele 077125, Romania

[4] Department of Engineering, Reykjavik University, Menntavegur 1, IS-102 Reykjavik, Iceland

E-mail: ioana@infim.ro and mihaela.florea@infim.ro

**Abstract.** Herein we present a comparative study among different mesoporous electron transporter layers (ETLs), namely nanometric m-TiO$_2$, m-SnO$_2$ and m-SnO$_2$ quantum dots (QDs), deposited by spray coating method. The experimental data correlated with the photovoltaic parameters indicate that the SnO$_2$ mesoporous layer obtained from the spray deposition of the in-house prepared QDs solution is the best candidate between the three used mesoporous ETLs. The use of the in-house prepared SnO$_2$ QDs solution presents smaller agglomerates composed of 3 nm NPs resulting in the formation of a thinner, more uniform, and compact mesoporous ETL, compared with the other two ETL solutions. The formamidinium-methylammonium-potassium (FAMA@10K) perovskite deposited on this m-SnO$_2$ QDs ETL presents a lower RMS, more uniformity and, a higher amount of PbI$_2$. Interestingly, this higher concentration for PbI$_2$ seems to enhance the performance of the perovskite solar cells (PSC), compared to the other two mesoporous ETLs. Our work reveals that SnO$_2$ QDs solution can be easily produced in the laboratory and it is more suited for the deposition of the mesoporous ETL when choosing a FAMA@10K configuration perovskite solar cell with power conversion efficiency (PCE) higher than 10%.

## 1. Introduction

Perovskite solar cells (PSCs) have demonstrated outstanding progress over less than a decade. The power conversion efficiency (PCE) has been improved, currently exceeding 25% [1], approaching the monocrystalline silicon solar cells and surpassing cadmium telluride (CdTe) and copper indium gallium selenide (CIGS) solar cells. The high efficiency of perovskite halide materials in solar cells is due to their unique properties. These include an adjustable band gap, strong light absorption, long carrier diffusion lengths, and good carrier motilities [2–5]. Such perovskite devices are comprised of an n-type layer (compact and mesoporous) deposited on a glass substrate covered with a thin transparent conductive oxide (ITO or FTO), a halide perovskite light absorber, a p-type hole transporter layer (HTL) and a metallic back electrode [6].

Exploration of the materials used for charge transporter layers represents a subject of high interest, due to their role in improving the performance and stability of PSCs. Most ETLs (electron transporter layers) are based on a combination of compact and mesoporous oxides, that have to be oxygen vacancy free which act as trap sites [7]. The compact layer's role is to block recombination between electrons in the transparent conductive oxide (TCO) layer and holes in the perovskite layer and must be uniform and dense to prevent short-circuits and equally distributed electric field, while the mesoporous layer should be porous and accommodate efficiently the perovskite layer in order to facilitate efficient collection and transport of electrons from the perovskite to the cathode.

Therefore, the ETL is considered one of the most important components of PSCs, where several morphological parameters, such as thickness, roughness, and particle size, or structural properties and density, play a key role in influencing the performance of PSCs. Most common ETLs used in PSCs are titanium dioxide ($TiO_2$), zinc oxide (ZnO), tin dioxide ($SnO_2$), fullerene ($C_60$), aluminum oxide ($Al_2O_3$), tungsten oxide ($WO_x$) and many other materials [8–11].

Among all the materials mentioned, $TiO_2$ is the most widely used in the architecture of mesoporous PSCs. It serves as both a compact and mesoporous ETL, owning a suitable band gap and high transmittance that ensure high performance of the PSCs [12–14]. However, $TiO_2$ presents certain disadvantages, like reduced electron mobility and optical instability [15]. $SnO_2$ represents an alternative to $TiO_2$ due to its unique properties such as high electron mobility (240 cm$^2$/Vs), excellent chemical stability, low photocatalytic activity, good anti-reflectivity, large band gap (3.6-4.2 eV) allowing a high stability under UV illumination, as well as an excellent band energy level alignment at the ETL/perovskite interface resulting in effective electron extraction and hole blocking [16]. In addition, $SnO_2$ quantum dots (QDs) films have been extensively investigated in PSCs due to their strong light scattering ability, fast electron transport and slow recombination, leading to the increase of PCE [17].

The transition of studying PSCs from small-scale areas to larger panels and eventual industrial-scale implementation represents the future trajectory of this technology. Spray coating is a deposition method used for inorganic thin layers used for different applications, and it is scalable for industrial use [18].

This study aims to find an easier, more accessible and scalable way to deposit the mesoporous ETL layers for halide perovskite solar cells. Thus, we make a comparative study of one $TiO_2$ and two $SnO_2$ mesoporous ETLs obtained by spray deposition of their corresponding colloidal solutions: a $TiO_2$ suspension obtained from a commercial paste, a $SnO_2$ commercial suspension and an in-house $SnO_2$ QDs suspension. To demonstrate the suitability of the in-house prepared $SnO_2$ QDs suspension for the preparation by spray coating of the mesoporous ETL we performed several investigations and compared them with the commercial one as well with the $TiO_2$ containing ETL layers. As a perovskite layer we are using $K_{0.1}FA_{0.7}MA_{0.2}PbI_{2.8}Cl_{0.2}$ (FAMA@10K) due to the promising results obtained on this composition [19].

## 2. Material and methods

All the precursors and solvents were used without any further purification and were purchased as follows: N,N-Dimethylformamide (DMF, Alfa Aesar 99.8%), dimethylsulfoxide (DMSO, Alfa Aesar 99.9%), chlorobenzene (CB, Sigma Aldrich 99.8%), isopropanol (IPA, Sigma Aldrich 98.5%), acetone (Sigma Aldrich), anhydrous ethanol (Sigma Aldrich 96%), 4-tert-butylpyridine (tBP, Sigma Aldrich 99.99%), acetonitrile (ACN, Sigma Aldrich 99.8%), titanium diisopropoxide bis(acetylacetonate) (Ti (iProp)2AcAc2, Sigma Aldrich 75 wt.% in isopropanol), lithium bis(trifluoromethanesulfonyl)imide (LiTFSI, Sigma Aldrich 99.99%), $TiO_2$ commercial paste (Ti-Nanoxide N/SP, Solaronix), Spiro-OMeTAD (Borun chemicals, 99.99%), lead chloride ($PbCl_2$, Sigma Aldrich 99.99%), lead iodide ($PbI_2$, Sigma Aldrich 99%), formamidinium iodide (FAI, Dyesol 99.9%), methylammonium iodide (MAI, Dyesol 99%), potassium iodide (KI, Merck 99.99%).

### 2.1. Device fabrication

The photovoltaic performance of the studied layers was probed by fabricating n-i-p devices with the following configurations: ITO/c-$TiO_2$/m-$TiO_2$, m-$SnO_2$ or $SnO_2$ QDs /FAMA@10K/Spiro-OMeTAD/Au and FTO/c-$TiO_2$ $SnO_2$ QDs /FAMA@10K/Spiro-OMeTAD/Au.

*2.1.1. Substrates*

Pre-cut indium tin oxide (ITO, 2.5 × 1.5 cm2) and fluorine tin oxide (FTO, 2.5 × 1.5 cm²) coated glass slides with a sheet resistance of $10\Omega/sq$ were purchased from Xin Yan Technology LTD. The substrates were washed with a detergent solution, and rinsed with distilled water and distilled water, acetone and isopropyl alcohol for 10 minutes each. An oxygen plasma treatment was applied for all the substrates for 15 minutes at 0.7 mbar.

*Compact $TiO_2$* - The compact layer of $TiO_2$ was fabricated by spray pyrolysis using an in-house automated apparatus [18]. A solution of Ti(iProp)2AcAc2 in isopropyl alcohol (1:30 volumetric ratio) was deposited through 30 spray sweeps keeping the nozzle at 15 cm height and a substrate temperature of 450 °C, using $N_2$ as carrier gas at $2kgf/cm^{-2}$ pressure and a sweep speed of $5000\ \mu m/s$. After deposition, the samples were treated at 450 °C for 30 minutes on the hot plate in air.

*Mesoporous TiO₂* - A commercial paste of Ti-Nanoxide T/SP from Solaronix which contains highly dispersed titania nano-particle paste for the deposition of transparent active mesoporous layers was used for the deposition of the mesoporous $TiO_2$ layer. The paste and the anhydrous ethanol were mixed in 1:100 (wt. ratio) to obtain a 1:100 dilutions for the suspension used at spray pyrolysis of the $TiO_2$ mesoporous layer. The suspension was dispersed on the $TiO_2$ compact layer following the steps used for the $TiO_2$ compact layer but at a substrate temperature of 120 °C. After deposition, the samples were treated at 120 °C for 10 minutes on the hot plate in the air.

*Mesoporous SnO₂* - A commercial suspension of $SnO_2$ colloidal solution 15% in water diluted with a mixture of ethanol/water (1:1 v/v) in order to obtain a dilution of 1:100 (wt. ratio) was deposited through 8 spray sweeps at 15 cm height and a substrate temperature of 180 °C, using $N_2$ as carrier gas at 2kgf/cm$^{-2}$ pressure and a sweep speed of 50000 $\mu m/s$. After deposition the samples were treated at the corresponding deposition temperature of 200 °C for 10 minutes on the hot plate for solvent evaporation. The m-$SnO_2$ films were then thermally treated at 500 °C in air with a heating ramp of 5 °C/min and kept for 1 hour at this temperature.

*SnO₂ QDs preparation* - A suspension of $SnO_2$ QDs was obtained following the method reported by Lu et al. [20]. This suspension was diluted to 3% and used to obtain a concentration of 1:100 (wt. ratio) $SnO_2$ QDs in a mixture of ethanol/water (1:1 v/v). The suspension was dispersed through 8 spray sweeps at 15 cm height and a substrate temperature of 200 °C, using $N_2$ as carrier gas at 2kgf/cm$^{-2}$ pressure and a sweep speed of 50000 $\mu m/s$. After deposition the samples were treated at the corresponding deposition temperature of 200 °C for 10 minutes on the hot plate for solvent evaporation. All the mesoporous films were then thermally treated at 500°C in air with a heating ramp of 5°C/min and kept for 1 hour at this temperature to ensure the fixation of the layer.

### 2.1.2. Perovskite solutions and films

The perovskite FAMA@10K solutions and films were prepared inside of a glove box, at room temperature (24 °C) at a concentration of 1.45 M. The FAMA@10K perovskite precursor solution was prepared by mixing MAI, FAI, KI, $PbI_2$, and $PbCl_2$ in DMF: DMSO (8.2:1) mixture. The resulting solution was stirred for 1 h at room temperature before being used. A volume of 50 $\mu L$ from the perovskite solution was deposited by spin coating at 2000 rpm for 50 s, and after 15 s from the start of the spin coating process, 100 $\mu L$ of CB was dropped in to assist perovskite crystallization. Perovskite films formation was completed through annealing on a hot plate at 140 °C for 3 minutes. This potassium-doped perovskite (FAMA@10K) was chosen because it was previously demonstrated by our group [19] that it is more stable than the undoped absorber $FA_{0.8}MA_{0.2}PbI_{2.8}Cl_{0.2}$ (FAMA).

### 2.1.3. Hole transport layer (HTL) and gold (Au) electrode

Spiro-OMeTAD solution (used as HTL) was prepared by dissolving spiro-OMeTAD (80 mg), 4-tert-butylpyridine (28 $\mu L$) and Li-TSFI salt solution (18 $\mu L$ taken from a solution of 520 mg/$\mu L$ in acetonitrile) in 1 $mL$ of CB. The Spiro-OMeTAD solution was deposited by spin coating at 3000 rpm for 30 s, over the perovskite layer. The samples were stored overnight in a glovebox at room temperature to dry. Finally, Au electrodes (100 nm) were deposited by magnetron sputtering through a shadow mask finalizing devices with an active area of 0.09 cm$^2$.

## 2.2. Characterization

X-ray Diffraction (XRD) patterns were recorded in symmetrical coplanar geometry, using a Bruker AXS (D8 ADVANCE), with a Cu anode as source and a Ni filter to eliminate the Cu-K$\beta$ contribution to the detected signal, with a dichromatic (Cu-K$\alpha$1 = 1.54060 ˚A and Cu-K$\alpha$2 = 1.544 ˚A) incident beam. The XRD data were analyzed using MAUD version 2.99 software to determine the average crystallite size and lattice parameters by Rietveld refinement.

High-Resolution Transmission Electron Microscopy (HRTEM) A JEOL 2100 Transmission Electron Microscope, equipped with High Resolution Polar Piece has been used for morpho-structural investigations.

Scanning Electron Microscopy (SEM) images were recorded using a Carl Zeiss Gemini 500 Field Emission Scanning Electron Microscope. The samples were imaged using an acceleration voltage of 5 kV (EHT), a working distance between 3 and 8 mm (WD) and different magnifications in high vacuum mode.

Atomic Force Microscopy (AFM) investigations were made using an atomic force microscope NT-MDT Aura Ntegra Prima in non-contact mode.

Device measurement protocol. Current-voltage characteristics were measured under simulated sunlight AM 1.5G from an Oriel VeraSol-2 Class AAA LED Solar Simulator, coupled with a Keithley 2400 source meter. The solar simulator was calibrated with an NREL certified KG5 filtered Si reference diode to an irradiation intensity of 100 mW/cm$^2$. Electrical measurements were performed in ambient environment, around 25 °C, starting just above the $V_{oc}$ to $J_{sc}$ and backward, at 20 mV/s, to avoid the sample pre-poling. All devices were measured in air without encapsulation. The open-circuit voltage decay was measured using an Oscilloscope PicoScope 444 measuring at 1 ms interval.

A dynamic J-V model analysis [21] is necessary to account for the hysteresis effects observed in current-voltage measurements of typical PSCs. By comparing experimental current density-voltage (J-V) measurements with the results of the dynamical model, the effects of the ETL can be identified by monitoring certain input parameters of the model (series resistance $R_s$, shunt resistance $R_{sh}$, thermal voltage $V_{th}$, ion capacitor parametrization, $C_0$, $C_1$ and recombination current parametrizations, $a$, $b$). Building upon the dynamical model developed in reference [21], we employed the Nelder-Mead (NM) algorithm [22] as implemented in the Python SciPy library to fit the results of the J-V dynamical model to the experimental measurements. Employing the NM algorithm, the input model parameters are iteratively adjusted to achieve a calculated J-V curve that closely aligns with the experimental values, using a target threshold for the coefficient of determination ($R^2$).

## 3. Results and discussion

In this study, we explored the effect of using three different ETLs, namely colloidal m-$TiO_2$, m-$SnO_2$ and m-$SnO_2$ QDs deposited by spray coating on c-$TiO_2$, on the performance of PSCs based on triple cation perovskite FAMA@10K absorbing layer. Perovskite solar cells have the following configurations: (Conductive layer)/ c-ETL/ m-ETL/ perovskite/ Spiro-OMeTAD/ Au. For the devices containing $SnO_2$-QDs two types of conductive substrates (TCO) were employed, indium tin oxide (ITO) and fluoride tin oxide (FTO) to study their influence on the crystallization of the perovskite layer. Also, the influence of the annealing temperature was studied.

In order to investigate the fabricated devices, three types of mesoporous bilayers (c-$TiO_2$/ m-$TiO_2$, c-$TiO_2$/ m-$SnO_2$ or c-$TiO_2$/m-$SnO_2$ QDs) at a concentration of 1:100 (wt ratio) were deposited on the substrates using spray coating deposition.

The mesoporous ETL layers have been characterized through powder X-ray diffraction (Figure 1a) and by Rietveld refinement analysis (data presented in Table S1 in Supplementary Information) and the following were observed: the XRD diffractogram of the m-$TiO_2$ layer presents the rutile like structure of $TiO_2$ (cif no. 1534781 from COD) with crystallites size of 20 nm. The X-ray diffraction pattern of the m-$SnO_2$ layer from commercial suspension presents diffraction lines characteristic to 15.74% $SnO_2$ rutile structure (cif. no. 1534781 from COD) with crystallite sizes of 21 nm, 60.64% $TiO_2$ anatase structure (cif. no. 1526931) with crystallite sizes of 35 nm, and to 23.62% ITO glass substrate (cif. no. 1010341). The X-ray diffraction pattern of the m-$SnO_2$ QDs layer shows diffraction lines characteristic to $SnO_2$ rutile structure (cif no. 1534781 from COD) with crystallite sizes of 19 nm. The powder XRD patterns and Rietveld refinement analysis of FAMA@10K perovskite films deposited on colloidal m-TiO2, m-$SnO_2$ and respectively, m-$SnO_2$ QDs revealed that all of them present diffraction lines at $2\theta$ = 14.01°, 19.94°, 24.38°, 28.2°, 31.65°, 34.8°characteristic to the black perovskite $FAPbI_3$ phase (in accordance with JCPDS no.00-069-1000), and diffraction lines characteristic to $PbI_2$ (cif. no. 1011333 from COD). However, there were some differences between the samples. Both perovskite and $PbI_2$ are present in different ratios and the crystallite size of FAMA@10K varies depending on the m-ETL used (Table 1).

The presence of residual PbI$_2$ suggests an incomplete transformation of the perovskite phase. Nevertheless, according to the literature, PbI$_2$ may be beneficial to passivate defects and decrease charge recombination in PSCs [23].

Table 1: Perovskite/ PbI$_2$ ratio and perovskite crystallite size in relation to the m-ETL used.

| m-ETL | FAMA@10K : PbI$_2$ ratio (%) | FAMA@10K crystallite size (nm) |
|---|---|---|
| m-TiO2 | 78.5 : 21.4 | 186.4 ± 2.2 |
| m-SnO$_2$ comm | 88.8 : 11.1 | 295.1 ± 12.5 |
| m-SnO$_2$ QDs | 79.1 : 20.8 | 212.0 ± 13.6 |

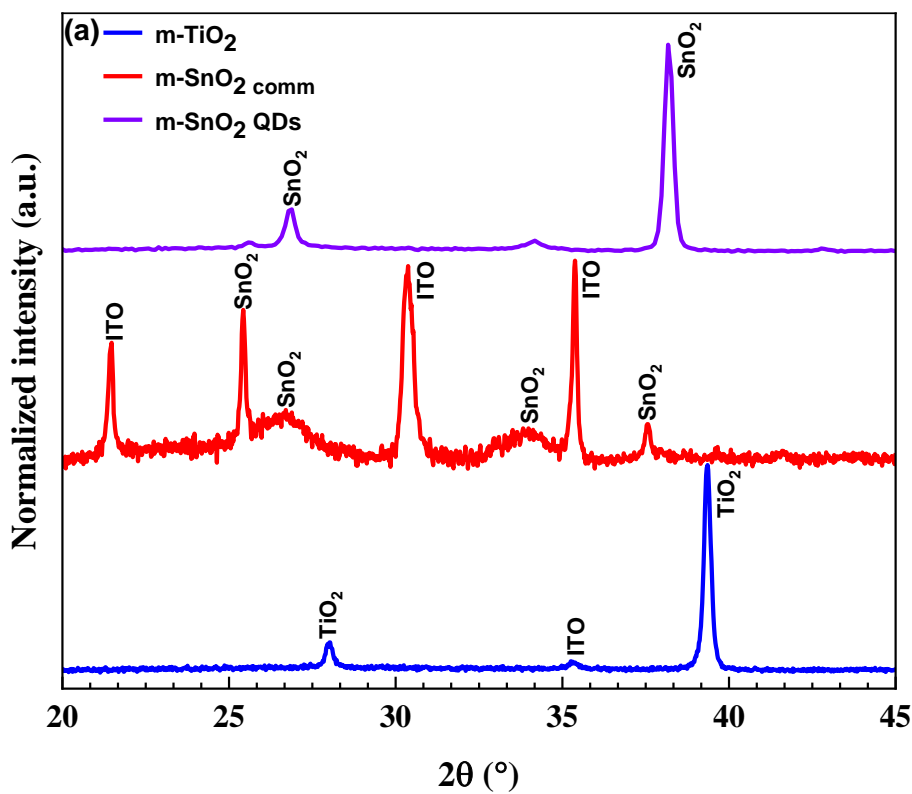

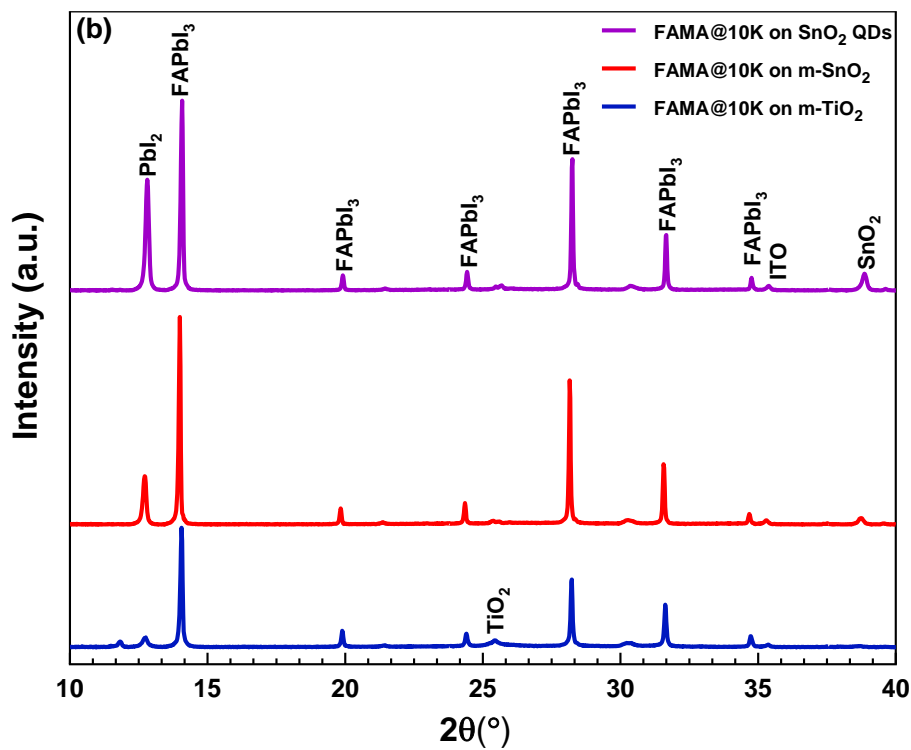

Figure 1: (a) XRD patterns of the ETL layers; (b) XRD patterns of the perovskite film FAMA@10K within the PSCs.

High-Resolution Transmission Electron Microscopy (HRTEM) imaging (Figure 1, a and b) obtained on the suspensions used for the deposition of the $SnO_2$ mesoporous layers revealed fine nanoparticles (NPs) with an expected quasi-spherical morphology. Selected Area Diffraction (SAED) proved the crystalline nature of $SnO_2$ NPs with tetragonal crystal lattice (Insets of Figure 2, a and b). HRTEM imaging obtained on the $TiO_2$ paste suspension revealed fine faceted nanoparticles, quasi-hexagonal with anatase-like crystalline structure (according to SAED, insets in Figure 2c), with a main NP size of $TiO_2$ of roughly 18 nm. The mean NP size of $SnO_2$ in both commercial and QDs suspensions is similar, roughly 3 nm. In both suspensions, the NPs tend to aggregate in large formations, but this tendency seems to be more pronounced in the case of commercially available $SnO_2$ suspension.

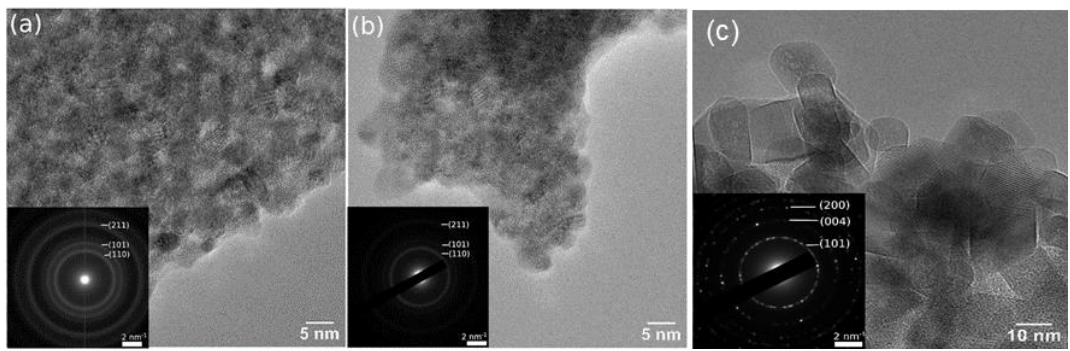

Figure 2: HRTEM images of the SnO₂ colloidal (a), QDs solution (b), and TiO₂ paste suspension (c). Inset: SAED

The topography of colloidal m-TiO2, m-SnO2, m-SnO2 QDs and respectively of the perovskite FAMA@10K deposited on the three types of ETLs over an area of 50 x 50 μm2 is shown in Figure 3. AFM results show compact films with full surface coverage, without pinholes for the perovskite.

The roughness values (RMS) for all of the ETL films are higher than other values reported in the literature [24] but the m-SnO₂ QDs present an RMS value lower than that of m-SnO₂ and for m-TiO₂. Even the roughness value of the perovskite FAMA@10K deposited on m-SnO₂-QDs is lower than the one deposited on m-SnO₂. The ETLs show different morphologies, root like surface for m-SnO₂, reticulated structure for m-TiO₂ and sphere-like structure for SnO₂ QDs. On the other hand, the RMS values reveal that the SnO₂ QDs present the smoothest film which is beneficial for a better basis for perovskite growth and also for forming better electronic contact with the ETL. [24] Due to a reduced formation of large clusters on the surface and reduced pinholes. The smooth surface of SnO₂ QDs leads to the formation of compact perovskite film, reducing the formation of large clusters on the surface and reducing pinholes. The same for the m-TiO₂, where the precursor solution fills its pores resulting in the formation of compact and very smooth film.

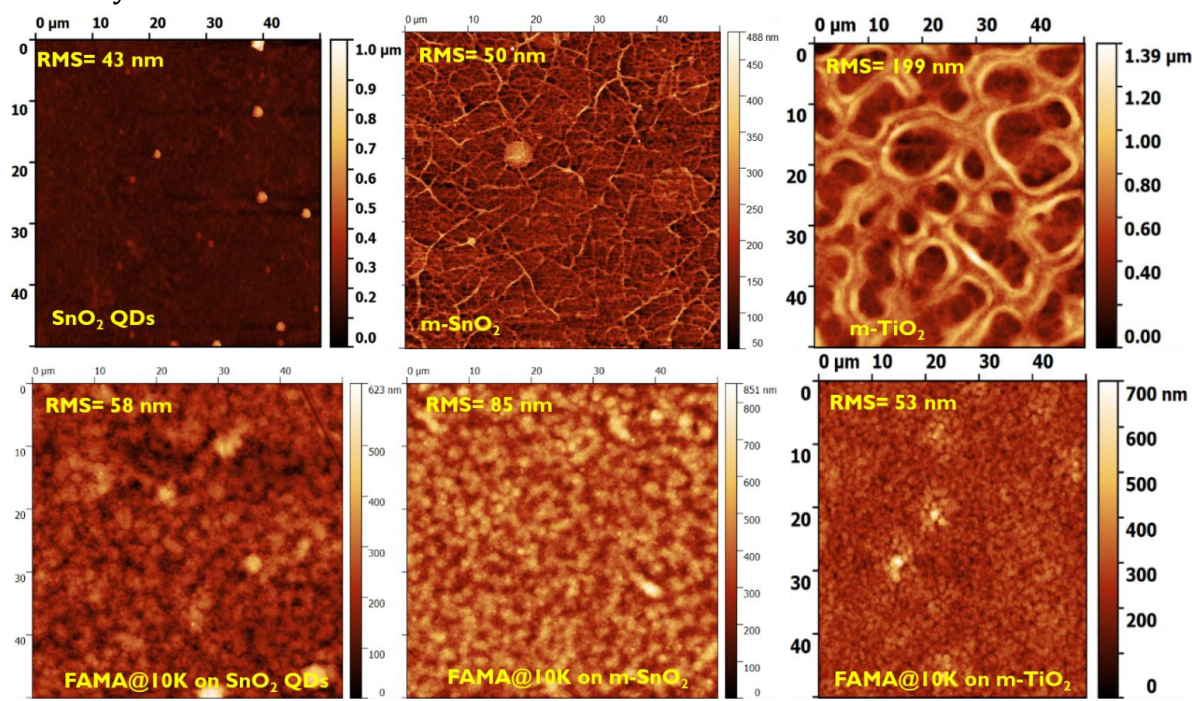

Figure 3: AFM images of the ETLs deposited from colloidal m-TiO₂, m-SnO₂ and m-SnO₂ QDs and of the FAMA@10K layers deposited on the different ETLs

The cross-section SEM images presented in Figure 4 show the thickness of the ITO substrate, ETL bilayers and of the perovskite in the two compositions. When comparing the characteristics of the three mesoporous ETL layers, it is very clear that m-SnO₂ QDs layer is more uniform and is thinner (~120 nm) than m-SnO₂ (~135 nm) and m-TiO₂ (~462 nm) layers.

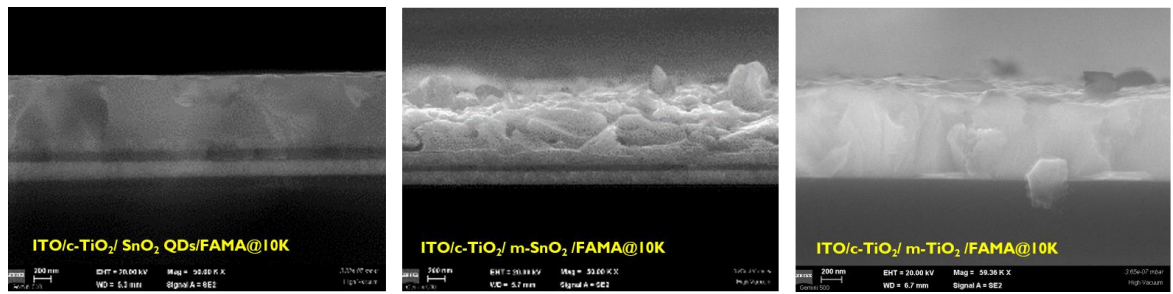

Figure 4: SEM images of FAMA@10K deposited on colloidal m-TiO₂, m-SnO₂ and m-SnO₂ QDs

In order to check the functionality and the photovoltaic characteristics, the fabricated solar devices with the following structure: ITO/c-TiO₂/ m-TiO₂ or m-SnO₂ or m-SnO₂ QDs/ FAMA@10K/Spiro-OMeTAD/Au have been investigated (Fig. 5) and the output parameters were extracted (Table 2).

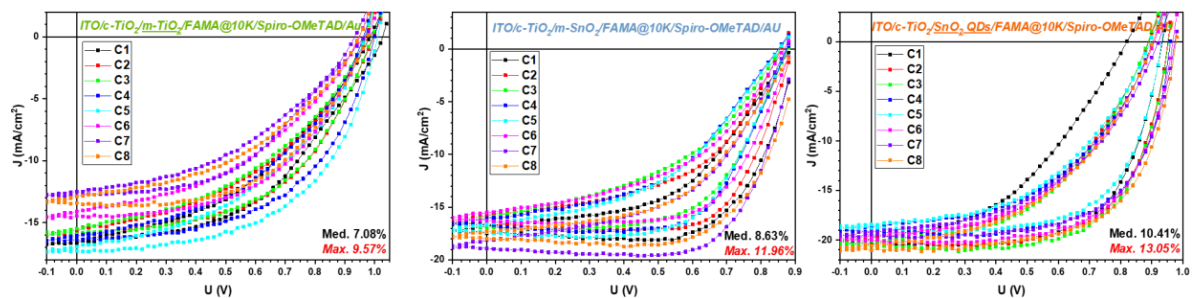

Figure 5: Forward (from -0.1 V to $V_{oc}$) and reverse (from $V_{oc}$ to -0.1 V) J-V curves of the champion cells

As shown in Figure 5, it can be easily noticed that both PSC with m-SnO₂ present big hysteresis and lower starting voltages compared to m-TiO₂, but even so, the best PCE value is obtained for m-SnO₂ QDs. If we discuss the values obtained for the J-V parameters of the champion PSCs presented in Table 2, we can observe that for the sample with m-SnO₂ QDs the values for $J_{sc}$, FF and $R_s$ are better, and, even though the $V_{oc}$ and $R_{sh}$ are lower than that of the sample containing m-TiO₂, it presents the highest PCE. The solar cell including m-TiO₂ present a PCE maximum of 9.57%, the one containing m-SnO₂ presents a PCE maximum of 11.96%, while the solar device using SnO₂ QDs exhibits the highest PCE of 13.05%. The highest performance of the device based on SnO₂ QDs is attributed mainly to the high-quality perovskite film with smooth surface. Most probably, the film efficiently reduce / suppress charge recombination at interfaces and enhance the transport of carriers in the solar devices. The low efficiency obtained using m-TiO$_2$ may be assigned to the presence of non-perovskite phase having poor optoelectronic properties as well as large band gap of about 2.48 eV.

Based on the characteristics of the mesoporous ETL layer and of the perovskite layer deposited on it, as well as the crystal size of the perovskite particles and the amount of PbI$_2$ compared to the other two ETLs systems, could be expected that the system having mSnO$_2$ QDs ETL presents higher performance.

From the average of the reverse and forward J-V scans of the champion PSC, we determined the series ($R_s$) and shunt ($R_{sh}$) resistances, as indicated in Table 2. However, this is just an estimation for the extracted quantities and a more rigorous calculation is performed using the dynamical J-V model [21], as described in the following.

Table 2: J-V parameters of champions PSCs with m-TiO$_2$, m-SnO$_2$, and m-SnO$_2$ QDs, using the averaged reverse-forward J-V characteristics.

| PSC Structure | $V_{oc}$ (V) | $J_{sc}$ (mA/cm$^2$) | FF (%) | $R_s$ ($\Omega$ cm$^2$) | $R_{sh}$ ($\Omega$ cm$^2$) | PCE (%) |
|---|---|---|---|---|---|---|
| ITO/c-TiO$_2$/m-TiO$_2$ [a] | 0.974 | 15.2 | 47 | 20 | 601 | 7.08 |
| ITO/c-TiO$_2$/m-SnO$_2$ [a] | 0.872 | 16.9 | 58 | 15 | 286 | 8.63 |
| ITO/c-TiO$_2$/m-SnO$_2$ QDs [a] | 0.929 | 19.6 | 57 | 14 | 438 | 10.41 |

[a] (1:100)/FAMA@10K/spiro-OMeTAD/Au; FF: fill factor; $J_{sc}$: short current density; $V_{oc}$: open-circuit voltage; $R_s$: series resistance; $R_{sh}$: shunt resistance.

From the set of measured J-V characteristics depicted in Figure 5, we selected three instances for each type of PSCs, which represent the minimum, maximum, and an intermediate efficiency case. We employed our dynamic hysteresis model [21] to determine the values of the relevant parameters for which the experimental J-V measurements and the calculated results are matched in a proportion of at least 90% in the R$^2$ coefficient of determination.

The curve fit representation shown in Figure 6 is achieved for all the systems, with different ETLs (TiO$_2$, SnO$_2$, SnO$_2$ QDs), proving the validity of the dynamical model. The JV curves present relatively large hysteresis, with a shift of the open circuit voltage between reverse and forward characteristics. We focused on a reduced set of model parameters [21] that provide the variations between the different classes of materials used as ETLs as well as for different samples within a class: the series resistance ($R_s$), the shunt resistance ($R_{sh}$), the thermal voltage ($V_{th}$), the parametrization of the ionic capacitance ($C_0$, $C_1$) and of the recombination current ($I_{rec0}$, $a$, $b$). In addition, a linear variation was considered for $R_s$ and $R_{sh}$ along the reverse-forward scans. The series resistances are decreasing in the sequence TiO$_2$ (25 $\Omega$cm$^2$) → SnO$_2$ (17 $\Omega$cm$^2$) → SnO$_2$ QDs (16 $\Omega$cm$^2$), found at the end of the forward scan. This is consistent with the trend observed on the average J-V characteristics in Table 2. The R$_{sh}$ values are in the same range of 200 $\Omega$ cm$^2$ to 600 $\Omega$cm$^2$, similar to the values obtained from the averaged J-V characteristics. The thermal voltage, $V_{th}$, which parametrizes the diode recombination current and, in effect, sets the open circuit voltage, has a small variation that reflects the $V_{oc}$ modifications in the measured data: 46 mV (TiO₂) → 40 mV (SnO₂) → 42 mV (SnO₂ QDs). The $b$ parameter relates the recombination current to the ionic current and controls the hysteresis magnitude. It presents a maximum for SnO₂ based PSCs, which have the largest hysteresis, while TiO₂ based PSCs have a minimum hysteresis: $b$=162 (TiO₂) → $b$ = 591 (SnO₂) → $b$ = 868 (SnO₂ QDs). The other parameters do not present significant variations that would enable a more detailed analysis of the three classes of PSCs.

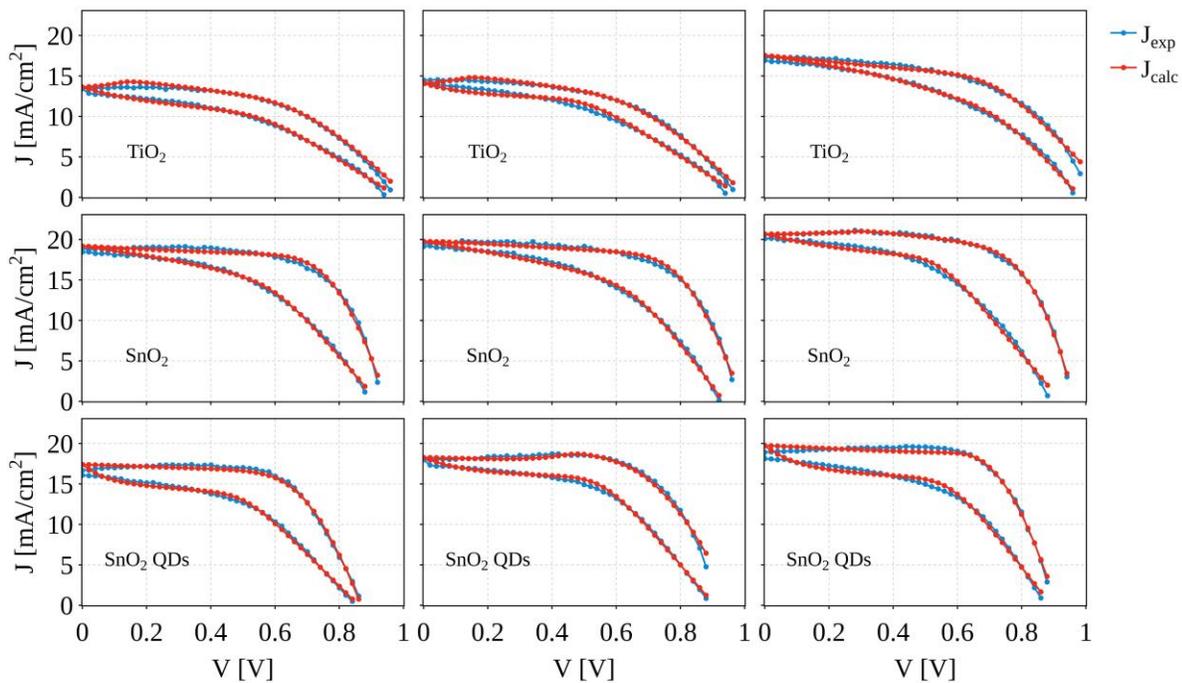

Figure 6: J-V curves fit of the experimental measurements (blue) using the dynamic JV model (red). Each row of plots contains the minimum, intermediate and maximum efficiency samples, for the three ETLs: TiO₂, SnO₂ and SnO₂ QDs.

Considering the results obtained, we further investigate the effect of two different deposition temperature of the SnO₂ QDs on the performance of PSCs. The XRD diffractograms of the perovskite deposited on SnO₂ QDs prepared at 120 °C and 200 °C (Figure 7) show that perovskite films exhibit the characteristic diffraction lines of tetragonal FAPbI₃. Besides the perovskite phase, both films contain also PbI₂, particularly in a significant amount in the film deposited on SnO₂ QDs prepared at 120 °C, which can lead to its degradation under illumination and humidity.

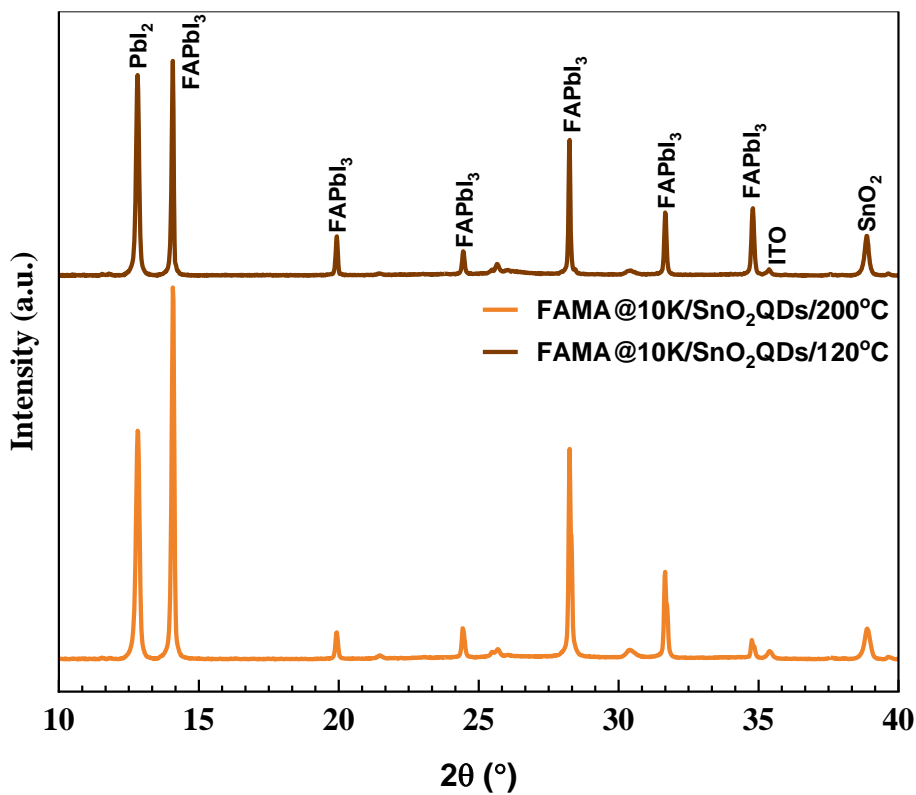

Figure 7: XRD diffractograms of the perovskite film FAMA@10K deposited on SnO₂ QDs prepared at 120 °C and 200 °C

On the other hand, AFM images of SnO₂ QDs prepared at different temperatures (Figure 8) reveal that the SnO₂ QDs are not uniformly distributed on the surface of the film prepared at 120 °C indicating that this temperature is not sufficient to grow the QDs. When the preparation temperature was increased at 200 °C, the surface became homogeneous and QDs were very well dispersed on the surface. The higher temperature needed for the tin oxide compared with the titanium oxide is due to the presence of water in the dispersions. Water is required in the solvent mixture in order to obtain stable dispersions. If only ethanol is used rapid sedimentation of both the commercial and in-house produced SnO₂ QDs occurs.

The RMS of the perovskite indicates that the perovskite film deposited on SnO₂ QDs annealed at 120 °C is rougher than the film deposited on SnO₂ QDs annealed at 200 °C. These results indicate that SnO₂ QDs deposited at different temperature impact the quality of the perovskite film that is deposited on it.

The cross-section SEM images of perovskite films deposited on SnO₂ QDs annealed at 120 °C and, respectively, at 200 °C (Figure 9) show the formation of uniform perovskite films in both cases. The effect of SnO₂ QD on the photovoltaic performance was subsequently explored. The device using SnO₂ QDs at 200 °C as ETL showed short circuit density ($J_{sc}$) of 19.6 mA/cm², an open-circuit voltage ($V_{oc}$) of 0.92 V and fill factor of 57 % corresponding to an overall conversion efficiency of 10.41% (Figure 10 A). Under the same conditions, the device with SnO₂ QDs at 120 °C as ETL demonstrated decreased $V_{oc}$ and FF values resulting in PCE of 7.39 % (Figure 10 B). This may be explained by the small amount of perovskite phase with a large amount of residual PbI₂ that has grown on top of a rough and non-uniform SnO₂ QDs layer, resulting in a reduced light absorption efficiency of the solar device.

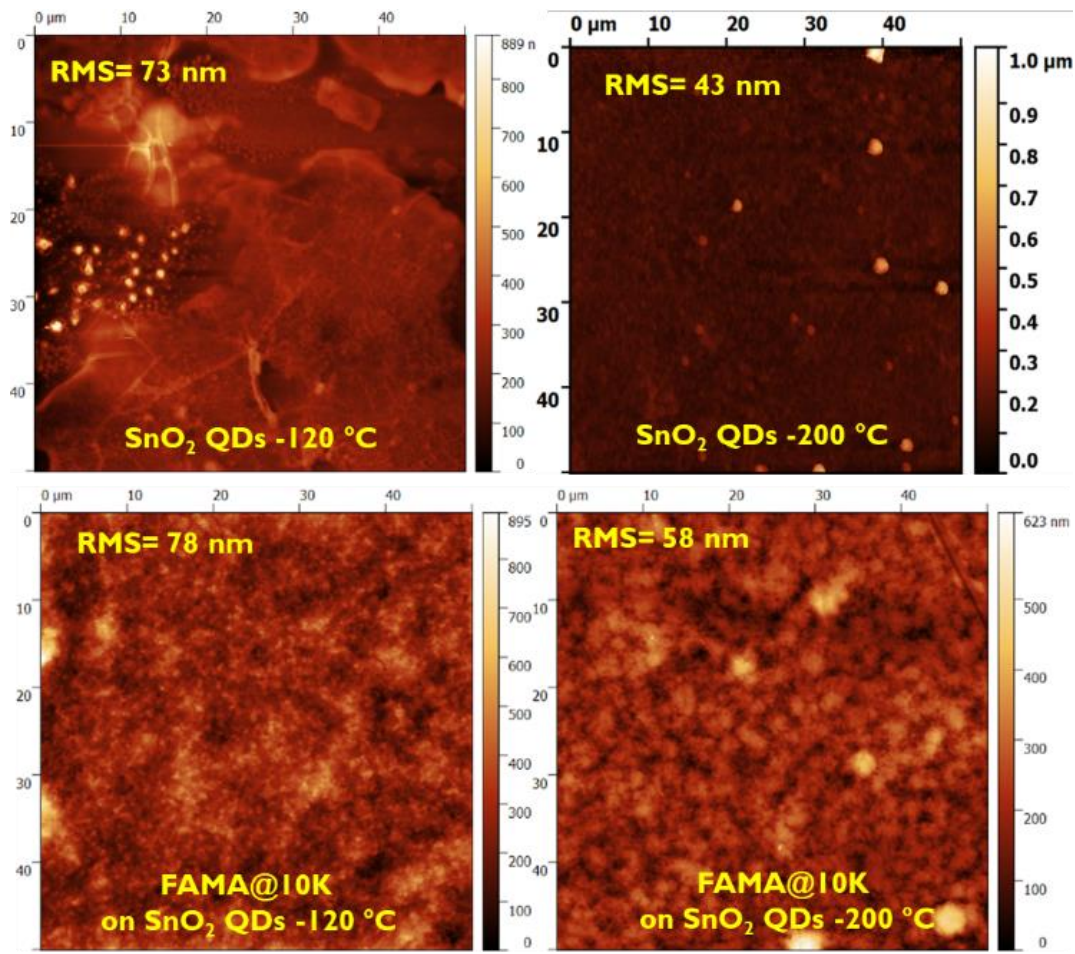

Figure 8: AFM images of FAMA@10K deposited on SnO₂ QDs prepared at 120 ˚C and 200 ˚C.

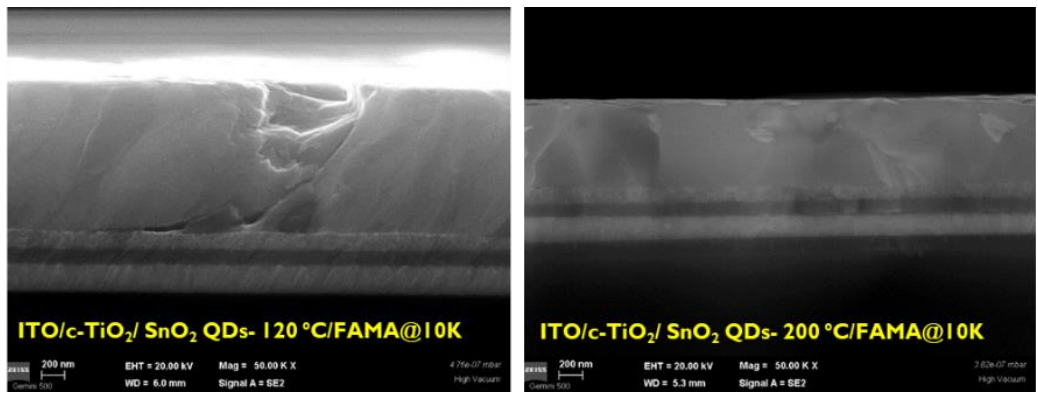

Figure 9: Cross SEM images of FAMA@10K deposited m-SnO₂ QDs prepared at 120 °C and 200 °C.

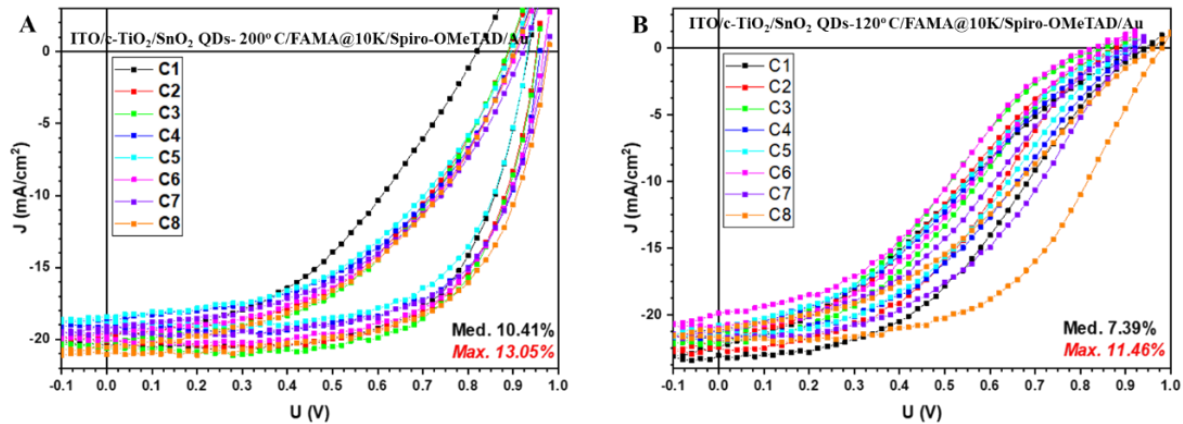

Figure 10: Forward and reverse J-V curves for PSCs using SnO₂ QDs annealed at 200 °C and 120 °C.

Considering the obtained results, we investigate the effect of the transparent conductive oxide substrate on the efficiency of PSCs. To date, ITO was found to be found to be superior to FTO in terms of sheet resistance, but it is more expensive than FTO substrates. In addition, FTO has been widely used in PSCs due to its high temperature resistance. Therefore, we will make a comparative study between using FTO and ITO.

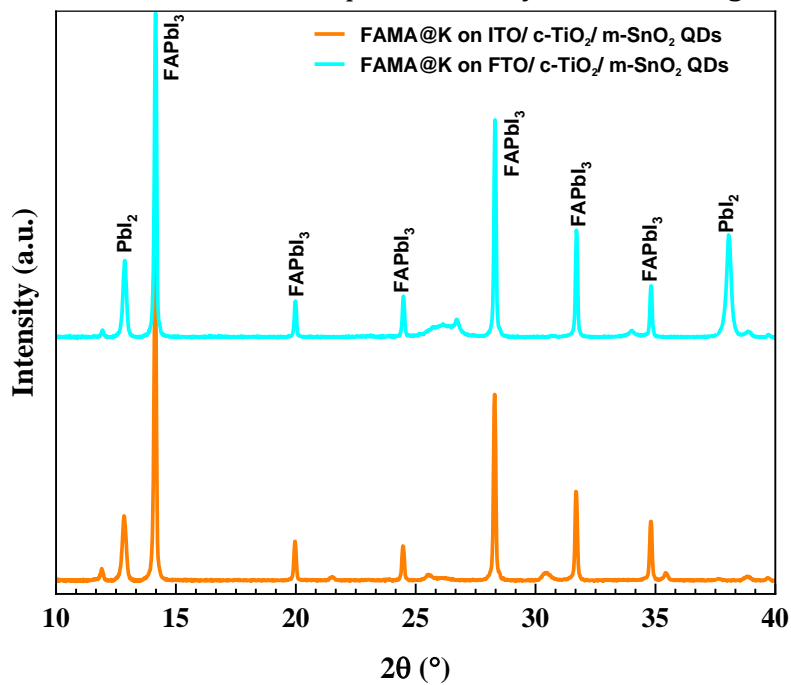

Figure 11: XRD diffractograms of the perovskite film FAMA@10K deposited on FTO/cTiO₂/ SnO₂ QDs or ITO/c-TiO₂/ SnO₂ QDs.

As we mentioned before, for the sake of comparison, we used two different substrates for the deposition of all layers c-TiO$_2$/m-SnO$_2$ QDs/ FAMA@10K/ spiro-oMeTAD/ Au, which are ITO and FTO, respectively. We studied by powder X-ray diffraction and Rietveld refinement analysis the FAMA@10K layers deposited considering the two different transparent electrodes (Figure 11).

We observed that both FAMA@10K layers contain the diffraction lines of the black perovskite and also of PbI$_2$ (Figure 11). The ratio between these two crystalline phases depends on the substrate used, when ITO was used, we obtained 86.5 % FAMA and 13.4 % PbI$_2$, but when FTO was used we obtained 83.1 % FAMA and 16.9 % PbI$_2$. Also, the crystallites size of FAMA were 175 nm when ITO was used and 190 nm when FTO was used. On the other hand, the roughness of perovskite film deposited on ITO was higher than the FAMA@10K deposited on FTO. As it can be observed in (Figure 12), the perovskite film deposited on m-SnO$_2$ QDs when ITO substrate was used (43.5 nm) presents a lower RMS value compared to when FTO was used (60 nm).

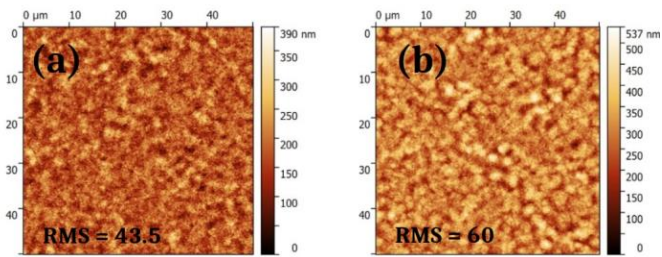

Figure 12: AFM images of FAMA@10K on: (a) ITO/c-TiO$_2$/ SnO$_2$ QDs or (b) FTO/cTiO$_2$/ SnO$_2$ QDs.

Moreover, the FAMA@10K film is non-uniform when ITO is used, in contrast with FTO for which the film is uniform. The cross section demonstrates uniform layers in both cases (Figure 13).

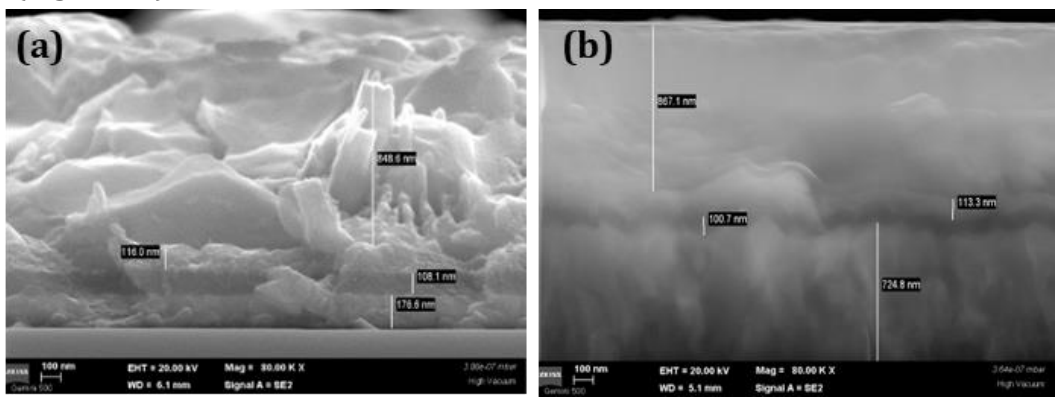

Figure 13: SEM cross-section of FAMA@10K on:(a) ITO/c-TiO$_2$/ SnO$_2$ QDs or (b) FTO/c-TiO$_2$/ SnO$_2$ QDs.

The photovoltaic parameters and conversion efficiency values obtained by using different TCO layer demonstrate that the PSCs with FTO-coated glass substrate achieved 14.90 % conversion efficiency as maximum PCE, which is higher than that obtained for PSCs with ITO coated glass substrate (Figure 14).

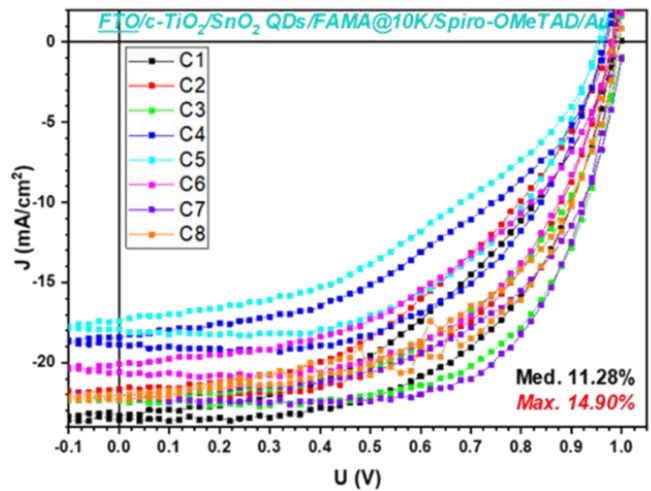

Figure 14: J-V curves for the champion cells measured under forward and reverse scan.

The higher PCE using FTO is attributed to higher short-circuit current ($J_{sc}$) and open-circuit voltage ($V_{oc}$) than those obtained for the one deposited on ITO. The reason for the higher $J_{sc}$ in the case of FTO is maybe attributed to the higher transmittance of FTO compared to ITO which could allow more light to be absorbed by the perovskite layer.

## 4. Conclusions

Three different mesoporous ETLs were used in halide perovskite solar cells and the photovoltaic device configuration is the following: TCO/c-$TiO_2$/m-$TiO_2$ or m-$SnO_2$ or m-$SnO_2$ QDs /FAMA@10K /Spiro-OMeTAD/Au. The mesoporous ETL layers were successfully deposited by an easy, more accessible and scalable method: the spray deposition. Following the same composition for the PSC, we prepared the mesoporous layer using a commercial $SnO_2$ suspension. The performance of the solar cells obtained with these three different ETL layers has been investigated. It was observed that the in-house prepared suspension presents smaller agglomerations composed of 3 nm NPs compared to the commercial one. The ETL layers prepared by the deposition of the two suspensions look similar at XRD but the Rietveld refinement analysis and AFM images indicate not only differences in the crystallite sizes and film composition but also a higher roughness for the one obtained from commercial suspension. Moreover, the perovskite surface seems to be smoother when deposited on the mesoporous ETL containing the in-house prepared $SnO_2$ QDs. When comparing the PCE and the J-V characteristics of the PSC obtained with the three mesoporous ETLs, better PCE and better J-V characteristics are achieved when $SnO_2$ QDs suspension is used for the mesoporous ETL. The J-V hysteresis and the trends concerning the variation of the PSC parameters are consistently explained by the dynamic electrical model. The study revealed that the optimal annealing temperature for $SnO_2$ QDs is 200 °C and a higher PCE is obtained when FTO is used. This study shows that in-house prepared $SnO_2$ QDs suspension is not only a good precursor for ETL deposition but a better choice than the commercial ones considering multiple factors.

## Acknowledgement

The research leading to these results has received funding from the EEA Grants 2014–2021, under Project contract no. 36/2021 (project code: EEA-RO-NO-2018-0106) and Core Program PN23080303. This work benefited from services and resources provided by EGI, with the dedicated support of CLOUDIFIN.

## Authors contribution

The manuscript was written by all authors. A.G.M. and I.D.V. contributed equally – QDs preparation, device fabrication, partial characterization, validation; S.D. – perovskite deposition, XRD and data interpretation; F.N. - validation and data interpretation; A.G.T. and C.B. – gold deposition, electrical measurements and data interpretation; M. E. – SEM analysis; A.C.K. – TEM analysis; A.C.I – Rietveld refinement; N.F. – theoretical calculation and data interpretation; M.C. – theoretical calculation and data interpretation; G.A.N. – theoretical calculation and data interpretation; A.M. – visualization, writing, review and editing; M.F. - conceptualization, writing, review, editing, I.P. – conceptualization, writing, review, editing and project administration.

## References

[1] NREL, Best Research-Cell Efficiency Chart, (2023). https://www.nrel.gov/pv/cell-efficiency.html.

[2] J.-H. Im, C.-R. Lee, J.-W. Lee, S.-W. Park, N.-G. Park, 6.5% efficient perovskite quantum-dot-sensitized solar cell, Nanoscale 3 (2011) 4088–4093. https://doi.org/10.1039/C1NR10867K.

[3] T.C. Sum, N. Mathews, Advancements in perovskite solar cells: photophysics behind the photovoltaics, Energy Environ. Sci. 7 (2014) 2518–2534. https://doi.org/10.1039/C4EE00673A.

[4] Q. Dong, Y. Fang, Y. Shao, P. Mulligan, J. Qiu, L. Cao, J. Huang, Electron-hole diffusion lengths > 175 μm in solution-grown CH3NH3PbI3 single crystals, Science (80-. ). 347 (2015) 967–970. https://doi.org/10.1126/science.aaa5760.

[5] Z. Zhang, M. Wang, L. Ren, K. Jin, Tunability of Band Gap and Photoluminescence in CH3NH3PbI3 Films by Anodized Aluminum Oxide Templates, Sci. Rep. 7 (2017)

1918. https://doi.org/10.1038/s41598-017-02144-x.

[6]   L. Xiong, M. Qin, C. Chen, J. Wen, G. Yang, Y. Guo, J. Ma, Q. Zhang, P. Qin, S. Li, G. Fang, Fully High-Temperature-Processed $SnO_2$ as Blocking Layer and Scaffold for Efficient, Stable, and Hysteresis-Free Mesoporous Perovskite Solar Cells, Adv. Funct. Mater. 28 (2018) 1706276. https://doi.org/https://doi.org/10.1002/adfm.201706276.

[7]   X.-X. Gao, Q.-Q. Ge, D.-J. Xue, J. Ding, J.-Y. Ma, Y.-X. Chen, B. Zhang, Y. Feng, L.-J. Wan, J.-S. Hu, Tuning the Fermi-level of $TiO_2$ mesoporous layer by lanthanum doping towards efficient perovskite solar cells, Nanoscale 8 (2016) 16881–16885. https://doi.org/10.1039/C6NR05917A.

[8]   A.G. Tomulescu, V. Stancu, C. Beşleagă, M. Enculescu, G.A. Nemneş, M. Florea, V. Dumitru, L. Pintilie, I. Pintilie, L. Leonat, Reticulated Mesoporous $TiO_2$ Scaffold, Fabricated by Spray Coating, for Large-Area Perovskite Solar Cells, Energy Technol. 8 (2020) 1900922. https://doi.org/https://doi.org/10.1002/ente.201900922.

[9]   J. Han, H. Kwon, E. Kim, D.-W. Kim, H.J. Son, D.H. Kim, Interfacial engineering of a ZnO electron transporting layer using self-assembled monolayers for high performance and stable perovskite solar cells, J. Mater. Chem. A 8 (2020) 2105–2113. https://doi.org/10.1039/C9TA12750J.

[10]  K. Wojciechowski, T. Leijtens, S. Siprova, C. Schlueter, M.T. Hörantner, J.T.-W. Wang, C.-Z. Li, A.K.-Y. Jen, T.-L. Lee, H.J. Snaith, C60 as an Efficient n-Type Compact Layer in Perovskite Solar Cells, J. Phys. Chem. Lett. 6 (2015) 2399–2405. https://doi.org/10.1021/acs.jpclett.5b00902.

[11]  D. Wang, Q. Chen, H. Mo, J. Jacobs, A. Thomas, Z. Liu, A bilayer $TiO_2/Al_2O_3$ as the mesoporous scaffold for enhanced air stability of ambient-processed perovskite solar cells, Mater. Adv. 1 (2020) 2057–2067. https://doi.org/10.1039/D0MA00562B.

[12]  D. Bi, C. Yi, J. Luo, J.-D. Décoppet, F. Zhang, S.M. Zakeeruddin, X. Li, A. Hagfeldt, M. Grätzel, Polymer-templated nucleation and crystal growth of perovskite films for solar cells with efficiency greater than 21%, Nat. Energy 1 (2016) 16142. https://doi.org/10.1038/nenergy.2016.142.

[13]  F. Wang, M. Yang, Y. Zhang, L. Yang, L. Fan, S. Lv, X. Liu, D. Han, J. Yang, Activating Old Materials with New Architecture: Boosting Performance of Perovskite Solar Cells with $H_2O$-Assisted Hierarchical Electron Transporting Layers, Adv. Sci. 6 (2019) 1801170. https://doi.org/https://doi.org/10.1002/advs.201801170.

[14]  M.F. Mohamad Noh, C.H. Teh, R. Daik, E.L. Lim, C.C. Yap, M.A. Ibrahim, N. Ahmad Ludin, A.R. bin Mohd Yusoff, J. Jang, M.A. Mat Teridi, The architecture of the electron transport layer for a perovskite solar cell, J. Mater. Chem. C 6 (2018) 682–712. https://doi.org/10.1039/C7TC04649A.

[15]  B. Roose, J.-P.C. Baena, K.C. Gödel, M. Graetzel, A. Hagfeldt, U. Steiner, A. Abate, Mesoporous $SnO_2$ electron selective contact enables UV-stable perovskite solar cells, Nano Energy 30 (2016) 517–522.

https://doi.org/https://doi.org/10.1016/j.nanoen.2016.10.055.

[16]   C. Guild, S. Biswas, Y. Meng, T. Jafari, A.M. Gaffney, S.L. Suib, Perspectives of spray pyrolysis for facile synthesis of catalysts and thin films: An introduction and summary of recent directions, Catal. Today 238 (2014) 87–94. https://doi.org/https://doi.org/10.1016/j.cattod.2014.03.056.

[17]   M. Kim, J. Jeong, H. Lu, T.K. Lee, F.T. Eickemeyer, Y. Liu, I.W. Choi, S.J. Choi, Y. Jo, H.-B. Kim, S.-I. Mo, Y.-K. Kim, H. Lee, N.G. An, S. Cho, W.R. Tress, S.M. Zakeeruddin, A. Hagfeldt, J.Y. Kim, M. Grätzel, D.S. Kim, Conformal quantum dot-SnO(2) layers as electron transporters for efficient perovskite solar cells., Science 375 (2022) 302–306. https://doi.org/10.1126/science.abh1885.

[18]   J. Leng, Z. Wang, J. Wang, H.-H. Wu, G. Yan, X. Li, H. Guo, Y. Liu, Q. Zhang, Z. Guo, Advances in nanostructures fabricated via spray pyrolysis and their applications in energy storage and conversion, Chem. Soc. Rev. 48 (2019) 3015–3072. https://doi.org/10.1039/C8CS00904J.

[19]   S. Derbali, K. Nouneh, M. Florea, L.N. Leonat, V. Stancu, A.G. Tomulescu, A.C. Galca, M. Secu, L. Pintilie, M.E. Touhami, Potassium-containing triple-cation mixed-halide perovskite materials: Toward efficient and stable solar cells, J. Alloys Compd. 858 (2021) 158335. https://doi.org/https://doi.org/10.1016/j.jallcom.2020.158335.

[20]   X. Lu, H. Wang, Z. Wang, Y. Jiang, D. Cao, G. Yang, Room-temperature synthesis of colloidal SnO2 quantum dot solution and ex-situ deposition on carbon nanotubes as anode materials for lithium ion batteries, J. Alloys Compd. 680 (2016) 109–115. https://doi.org/https://doi.org/10.1016/j.jallcom.2016.04.128.

[21]   N. Filipoiu, A.T. Preda, D.-V. Anghel, R. Patru, R.E. Brophy, M. Kateb, C. Besleaga, A.G. Tomulescu, I. Pintilie, A. Manolescu, G.A. Nemnes, Capacitive and Inductive Effects in Perovskite Solar Cells: The Different Roles of Ionic Current and Ionic Charge Accumulation, Phys. Rev. Appl. 18 (2022) 64087. https://doi.org/10.1103/PhysRevApplied.18.064087.

[22]   F. Gao, L. Han, Implementing the Nelder-Mead simplex algorithm with adaptive parameters, Comput. Optim. Appl. 51 (2012) 259–277. https://doi.org/10.1007/s10589-010-9329-3.

[23]   B. Park, N. Kedem, M. Kulbak, D.Y. Lee, W.S. Yang, N.J. Jeon, J. Seo, G. Kim, K.J. Kim, T.J. Shin, G. Hodes, D. Cahen, S. Il Seok, Understanding how excess lead iodide precursor improves halide perovskite solar cell performance, Nat. Commun. 9 (2018) 3301. https://doi.org/10.1038/s41467-018-05583-w.

[24]   V.P. Hoang Huy, C.-W. Bark, Review on Surface Modification of SnO2 Electron Transport Layer for High-Efficiency Perovskite Solar Cells, Appl. Sci. 13 (2023). https://doi.org/10.3390/app131910715.